# Magnetic Field-Enhanced Oxygen Reduction Reaction for Electrochemical Hydrogen Peroxide Production with Different Cerium Oxide Nanostructures

Caio Machado Fernandes[1], Aila O. Santos[2], Vanessa S. Antonin[1], João Paulo C. Moura[1], Aline B. Trench[1], Odivaldo C. Alves[2], Yutao Xing[3], Júlio César M. Silva[2], Mauro C. Santos[1*]

[1]*Laboratório de Eletroquímica e Materiais Nanoestruturados, Centro de Ciências Naturais e Humanas, Universidade Federal do ABC – Santo André, SP – Brasil*

[2]*Laboratório de Materiais da UFF, Instituto de Química, Universidade Federal Fluminense – Niterói, RJ – Brasil*

[3]*Laboratório de Microscopia Eletrônica de Alta Resolução, Centro de Caracterização Avançada para a Indústria de Petróleo, Universidade Federal Fluminense – Niterói, RJ – Brasil*

Corresponding author: mauro.santos@ufabc.edu.br

## Abstract

We investigated cerium oxide nanoparticles of various morphologies (nanosheets, nanocubes, and nanoparticles) supported on carbon Vulcan XC-72 for the two-electron oxygen reduction reaction (ORR). It was used a continuous magnetic field (2000 Oe) for the first time in the literature. The best results were for 5% (w/w) $CeO_2$ for all three different morphologies, more than doubling the ring current, enhancing the hydrogen peroxide selectivity from 51% (Vulcan XC-72) to 84-89%, and modifying the onset potential to lesser negative values. The presence of the magnetic field led to even higher ring currents with 5% (w/w) $CeO_2$, $H_2O_2$ selectivity from 54% (Vulcan XC-72) to 88-96% and changing even more the onset potential. Those results were correlated with the Zeeman effect, the Lorentz force, generating magnetohydrodynamic effects, the Kelvin force, and the formation of Bound Magnetic Polarons. This pioneering research introduces an innovative approach, highlighting the potential of an external continuous magnetic field.

## 1. Introduction

In recent times, the quest for sustainable and environmentally benign chemical processes has driven significant interest in the field of electrochemical reactions. Among these reactions, the Oxygen Reduction Reaction (ORR) holds particular significance due to its potential role in revolutionizing hydrogen peroxide ($H_2O_2$) production methodologies. Traditional methods for $H_2O_2$ synthesis have long relied upon energy-intensive processes that exhibit unfavorable environmental impacts. Still, compounding this problem is the prominent concern surrounding the storage and transportation of $H_2O_2$, which is marred by inherent instability and safety risks. As a result, the pressing need for cleaner and more energy-efficient alternatives has prompted an exploration of electrochemical approaches for ORR-based hydrogen peroxide generation to address both the energy-intensive nature of conventional $H_2O_2$ production and the challenges associated with its safe handling and distribution [1-4].

The two-electron pathway for the Oxygen Reduction Reaction (ORR) offers an environmentally sustainable, cost-effective avenue for electrogenerating $H_2O_2$. This approach minimizes the reagent consumption inherent in traditional synthesis while operating in solvent-free environments. Moreover, it enables in-situ hydrogen peroxide generation, streamlines product logistics, and enhances ease of storage and handling [5-7].

This research effort, the foremost hurdle faced by scientific groups, refers to the discovery of catalysts boasting elevated electrocatalytic efficacy coupled with selectivity, durability, and steadfast stability. Such catalysts are indispensable for seamless translation to larger-scale applications, specifically in the context of flow cells, wherein optimization of procedures is paramount. Addressing this challenge requires an intricate understanding of diverse catalyst types and their performance characteristics, ranging from transition metals and alloys to innovative nanoscale materials [8-10].

Cerium oxide ($CeO_2$) nanostructures supported on carbon materials have emerged as promising candidates for advancing the two-electron Oxygen Reduction Reaction (ORR) due to their unique catalytic properties. These nanomaterials exhibit high surface area, redox activity, and oxygen vacancy formation, facilitating efficient electrocatalysis. By harnessing the intrinsic ability of cerium oxide to toggle between different oxidation states, these nanomaterials enable the promotion of the desired two-electron pathway while mitigating unwanted side reactions.

This distinctive capability presents a pathway to enhance the selectivity, efficiency, and durability of ORR catalysts, offering a sustainable avenue for electrochemical $H_2O_2$ production and other pertinent applications [11-15]. Indeed, our research group published in 2018 an article discussing ceria high aspect ratio nanostructures supported on Carbon for hydrogen peroxide electrogeneration, proving the efficiency of this material for such purpose [16].

Persistent efforts to enhance $H_2O_2$ electrogeneration continue through structural and compositional alterations within electrocatalysts. However, despite extensive exploration of defect engineering, phase transitions, and doping, progress in electrocatalytic reactions remains constrained by conventional catalyst design and modification. A pioneering and promising solution, magnetic field-enhanced electrocatalysis, has arisen as an innovative avenue for augmenting electrochemical reactions. This novel strategy harnesses magnetic fields to offer an advanced and highly prospective approach to catalytic enhancement [17, 18].

Experimental observations have consistently highlighted the constructive impact of direct and indirect magnetic phenomena across a diverse spectrum of electrochemical reactions [19-27]. Within the context of Oxygen Reduction Reactions (ORR), the magnetic field-induced influences are notably characterized by magnetohydrodynamic effects stemming from the interplay of the Lorentz force ($F_L$) and the Kelvin force ($F_K$). Notably, the second one exerts a pronounced influence on electrochemical processes by facilitating the enhanced mass transfer of magnetic species. This heightened mass transfer dynamically amplifies the reaction kinetics in immediate proximity to the electrode, accelerating the reaction rate [28].

Following this thought, cerium oxide nanoparticles with different morphologies (nanosheets, nanocubes, and nanoparticles – all present ferromagnetism at room temperature) supported on carbon Vulcan XC-72 were evaluated as electrocatalysts for two-electrons ORR The magnetic behavior of $CeO_2$ enhances the kinetics and efficiency of electron transfer reactions when subjected to a continuous magnetic field.

This groundbreaking and pioneering research introduces a potential innovation in traditional electrocatalysis, demonstrating that applying an external magnetic field on metal oxides can yield significant benefits.

## 2. Experimental Procedure
### 2.1. CeO$_2$ nanostructures synthesis
#### 2.1.1. CeO$_2$ nanocubes

Cubic ceria (CeO$_2$ NC) was synthesized hydrothermally using Ce(NO$_3$)$_3$·6H$_2$O as the metal precursor. A dispersion of 0.05 mol L$^{-1}$ Ce(NO$_3$)$_3$·6H$_2$O and 6.00 mol L$^{-1}$ NaOH in 53.45 mL was stirred magnetically until a light purple colloidal dispersion formed. The dispersion was then hydrothermally treated at 180°C for 24 h in a Teflon-lined stainless steel autoclave. After natural cooling, the resulting light yellow precipitate was separated by centrifugation, washed with distilled water (2 times) and ethanol (3 times), and air-dried at 85°C for 24 h. The final powdered product was collected for further characterization.

#### 2.1.2. CeO$_2$ nanosheets

A hydrothermal method was processed to synthesize ceria hexagonal nanosheets (CeO$_2$ NS) using Ce(NO$_3$)$_3$·6H$_2$O as the precursor. A dispersion containing Ce(NO$_3$)$_3$·6H$_2$O 0.03 mol L$^{-1}$ and NH$_4$OH 0.016 mol L$^{-1}$ was prepared under magnetic stirring for 20 min at room temperature. Then, the dispersion was transferred to a Teflon-lined stainless steel autoclave placed in an oven at 220 °C for 24 h. The autoclave was cooled to room temperature, and the residue was washed with water and ethanol two and three times, respectively, and dried in air at 85 °C for 24 h.

#### 2.1.3. CeO$_2$ nanoparticles

Cerium oxide nanoparticles (CeO$_2$ NP) were synthesized through the utilization of cerium nitrate as the precursor. A measured aliquot of trisodium phosphate solution (20 mL) at a concentration of 0.02 moles was meticulously introduced drop by drop into a solution of cerium nitrate (60 mL) containing 0.1 moles, employing continuous agitation. The ensuing reaction, conducted under these consistent conditions, spanned 30 min, yielding a white colloidal solution. Subsequently, the residual mixture was carefully transferred to a suitably designed autoclave vessel, wherein the hydrothermal treatment was executed at a temperature of 180°C for 15 h. Upon the culmination of the reaction, the autoclave vessel was allowed to attain ambient temperature. The resultant colloid was then meticulously isolated through centrifugation, following which the acquired product underwent successive purification steps involving thorough washing with doubly distilled water and ethanol, culminating in a final phase of drying at 80°C for 12 h.

### 2.2. Electrocatalysts preparation

The preparation of $CeO_2$ electrocatalysts, encompassing nanocubes, nanosheets, and nanoparticles, was undertaken in conjunction with carbon Vulcan XC-72 as a support matrix, featuring varying loadings (1%, 3%, 5%, and 10% w/w) of the metal oxide. This was achieved through the wet impregnation method. The method involved weighing the requisite quantity of $CeO_2$, which was subsequently mixed with 0.5 g of Vulcan XC-72 and suspended in 3 mL of deionized water. This amalgam was subjected to continuous magnetic stirring for 4 h. A sample of pure Carbon was also meticulously prepared as a comparative benchmark. Following this stage, the synthesized electrocatalysts underwent drying in an electric oven at a controlled temperature of 100 °C.

### 2.3. Physicochemical characterization

X-ray diffraction (XRD) analyses were conducted using a Panalytical model X'Pert Pro-PW3042/10 diffractometer equipped with Cu Kα radiation (λ = 0.1540 nm) at 40 kV and 40 mA. A solid-state X-Celerator detector was employed. Scans were conducted over a 2θ range of 20 to 80 °C at a scanning rate of 0.025 ° $s^{-1}$.

The nanostructures' scanning electron microscopy (SEM) analysis was performed using a JEOL JSM-7100 F SEM. The transmission electron microscopy (TEM) images were obtained using a JEOL JEM 2100F. The energy-dispersive X-ray spectroscopy (EDS) was used to generate elementary maps in scanning transmission electron microscopy (STEM) mode. Electrocatalyst contact angles were measured using a goniometer (G.B.X. Digidrop) to assess wettability.

X-ray photoelectron spectroscopy (XPS) was conducted using a Scienta Omicron ESCA+ spectrometer from Germany, utilizing monochromatic Al Kα (1486.7 eV) radiation. Shirley's method was employed for C 1s and O 1s high-resolution core-level spectra to remove the inelastic background. Spectral fitting, performed with CasaXPS software, involved unconstrained multiple Voigt profile fitting.

Magnetic measurements at room temperature were carried out using a physical property measurement system (PPMS), specifically the VersaLab by Quantum Design.

### 2.4. Oxygen Reduction Reaction (ORR)

Electrochemical experiments were conducted using a potentiostat/galvanostat (Autolab PGSTAT 302N) coupled with a rotating ring disc electrode system (RRDE) from Pine Instruments. The working electrode setup encompassed a glass carbon (GC) disc (0.2475 cm²) paired with a platinum ring (0.1866 cm²) at a collection factor of N = 0.26. In contrast, a 2 cm² Pt counter electrode and Hg|HgO reference electrode were utilized. The supporting electrolyte was 1 mol L$^{-1}$ NaOH. The deposition of electrocatalysts onto the GC disc was accomplished through drop casting. Homogenized dispersions of the electrocatalyst in water (2 mg mL$^{-1}$) were prepared via ultrasonication, and 20 μL of the resulting mixture was applied to the disc electrode surface. Following drying, a 20 μL aliquot of a 1:100 Nafion solution (v/v, Nafion: deionized water) was added to the electrode film and dried. The electrolyte was oxygen-saturated for 30 min before all electrochemical analyses, maintaining consistent flow during duplicate measurements conducted at room temperature with a scan rate of 5 mV s$^{-1}$. That procedure was performed in the absence and presence of a magnet capable of generating a magnetic field of 2000 Oe in the distance used (1.0 cm from the electrode). Those measurements were performed in triplicates and in alkaline media because that electrolyte guarantees more pronounced currents with less noise disturbance in RRDE experiments.

The oxygen reduction reactions in alkaline media are as following [29]:

$$O_2 + 2H_2O + 4e^- \rightarrow 4OH^-$$

$$O_2 + H_2O + 2e^- \rightarrow HO_2^- + OH^-$$

$$HO_2^- + H_2O + 2e^- \rightarrow 3OH^-$$

## 3. Results and discussion

### 3.1. Nanostructures characterization

The XRD patterns of the different CeO$_2$ nanostructures are displayed in Fig. 1. The structure is Fluorite#CaF2, with crystal class m-3m, a crystal cubic system, and space group Fm-3m (225), according to ICSD card no. 24887 [30]. The nanoparticles, nanosheets, and nanocubes all have the same characteristic peaks, observed at around 29, 33, 48, 56, 59, 70, 77, 79, and 88 degrees, confirming the success of the synthesis of ceria nanomaterials.

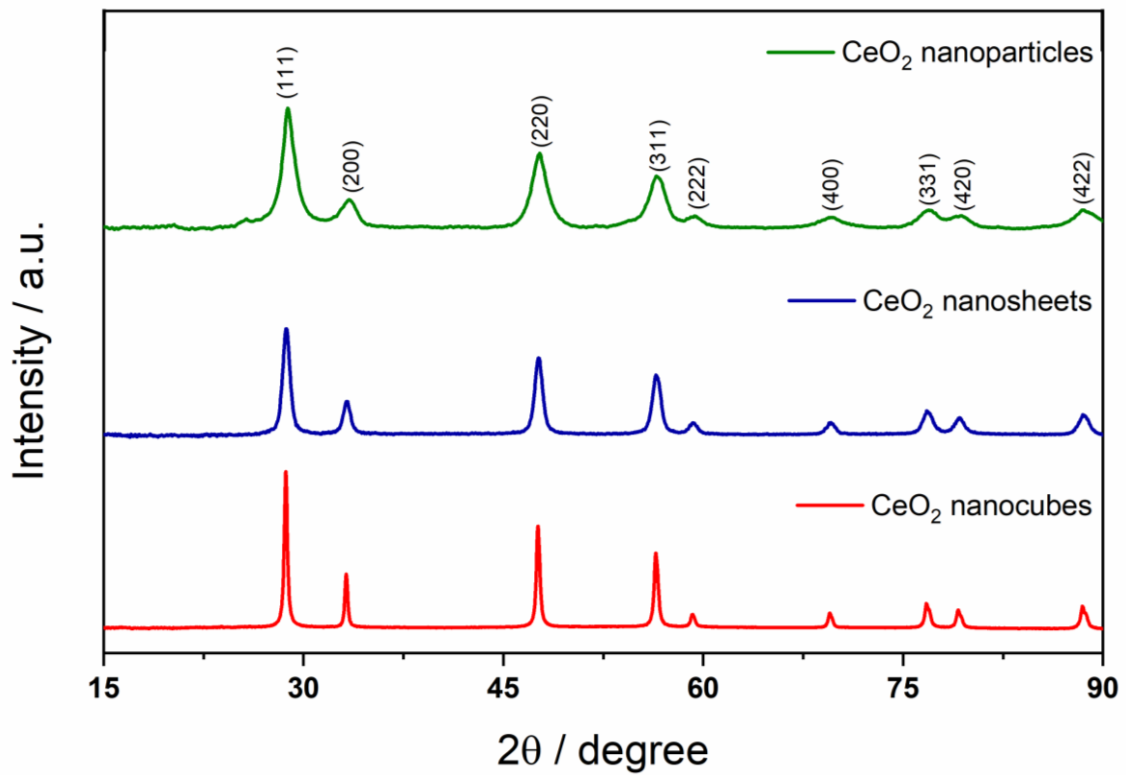

**Figure 1.** XRD patterns of different CeO$_2$ nanostructures.

The magnetization as a function of external magnetic field (M x H) were measured at 300 K and the results are shown in Fig. 2. All the three samples present ferromagnetic behavior associated with the presence of oxygen vacancies and Ce$^{3+}$ in the surface of the nanomaterial. The values of saturation magnetization, M$_s$, and coercive field, H$_c$, are in the same order as previous works and indicate weak nanoparticle ferromagnetism [31]. The distinct nanoparticle morphologies present different exposed crystalline planes with proper energy for vacancy formation that follows the order (110) < (100) < (111) [32]. The nanocubes present (100) as exposed plane and larger $M_s$ values than nanosheets with (110) as exposed plane and nanoparticles with no defined exposed plane [31].

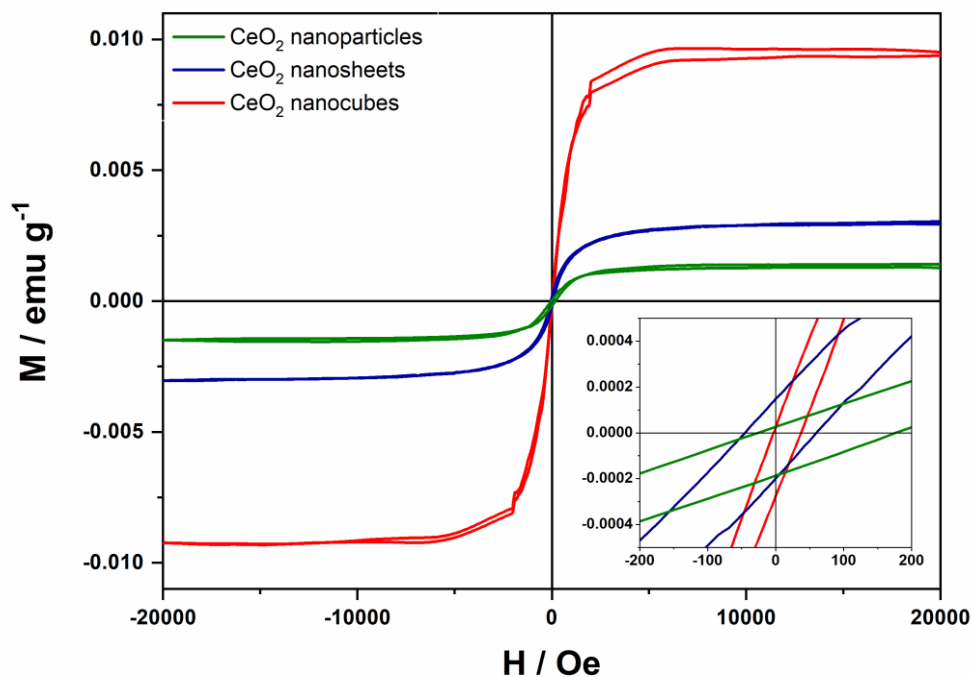

**Figure 2.** Magnetization as a function of external magnetic field of different $CeO_2$ nanostructures measured at 300 K. Insert shows the low field region.

A comprehensive investigation of the electrocatalyst morphology was conducted through SEM and TEM. The obtained SEM images (Fig. S1) vividly illustrate the uniform distribution of cerium oxide throughout the carbonaceous matrix, a pivotal characteristic for electrocatalysts of this category. Further insights were gained through TEM and High-Resolution TEM (HRTEM)(Fig. S2). These images unequivocally confirm the distinct shapes of ceria material: nanocubes depicted in Fig. S2 a and b, nanospheres and nanowires displayed in Fig. S2 c and d (termed as nanoparticles due to the variety of shapes observed), and hexagonal nanosheets showcased in Fig. S2 e and f. The dimensions of these ceria nanostructures were determined to be approximately 15-20 nm. This extensive characterization reaffirms the pervasive presence of $CeO_2$, underscoring its thorough dispersion within the Carbon Vulcan XC 72 substrate. Also, Fig. S3 shows the specific morphology of the materials obtained by controlled synthesis (nanocubes and nanosheets) without the carbon matrix.

Fig. 3 presents the bright-field STEM images and EDS elemental mapping results. These findings unequivocally demonstrate that the nanostructures comprise cerium (Ce) and oxygen (O). The data unquestionably establish the presence of cerium oxide material, confirming the distinct morphologies and its homogeneous dispersion within the carbon Vulcan XC72 matrix for all three tested materials.

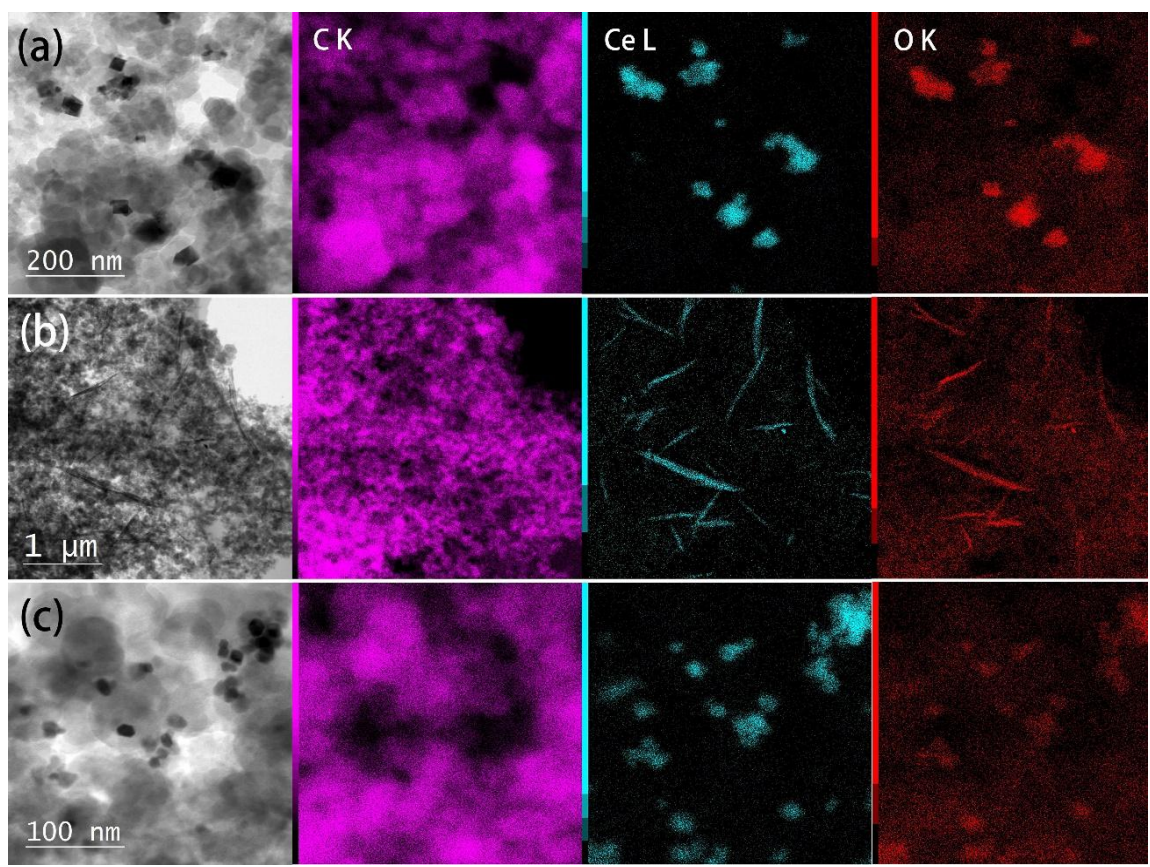

**Figure 3.** Bright-field STEM images and EDS elemental mapping results for 5% (a) $CeO_2$ nanocubes/C, (b) $CeO_2$ nanoparticles /C, and (c) $CeO_2$ nanosheets/C.

High-resolution XPS C 1s core-level spectra of Vulcan XC72, 5% $CeO_2$ NPs/C, 5% $CeO_2$ NCs/C, and 5% $CeO_2$ NSs/C samples are depicted in Fig. 4. These spectra were subjected to deconvolution into four components to explore the presence of distinct functional groups. The peak at approximately 284 eV was attributed to C-C groups, the peak at about 285 eV was designated for C-OH groups, the peak at around 286 eV was assigned to C=O groups, and the peak at approximately 289 eV was associated with -COOH groups [33, 34].

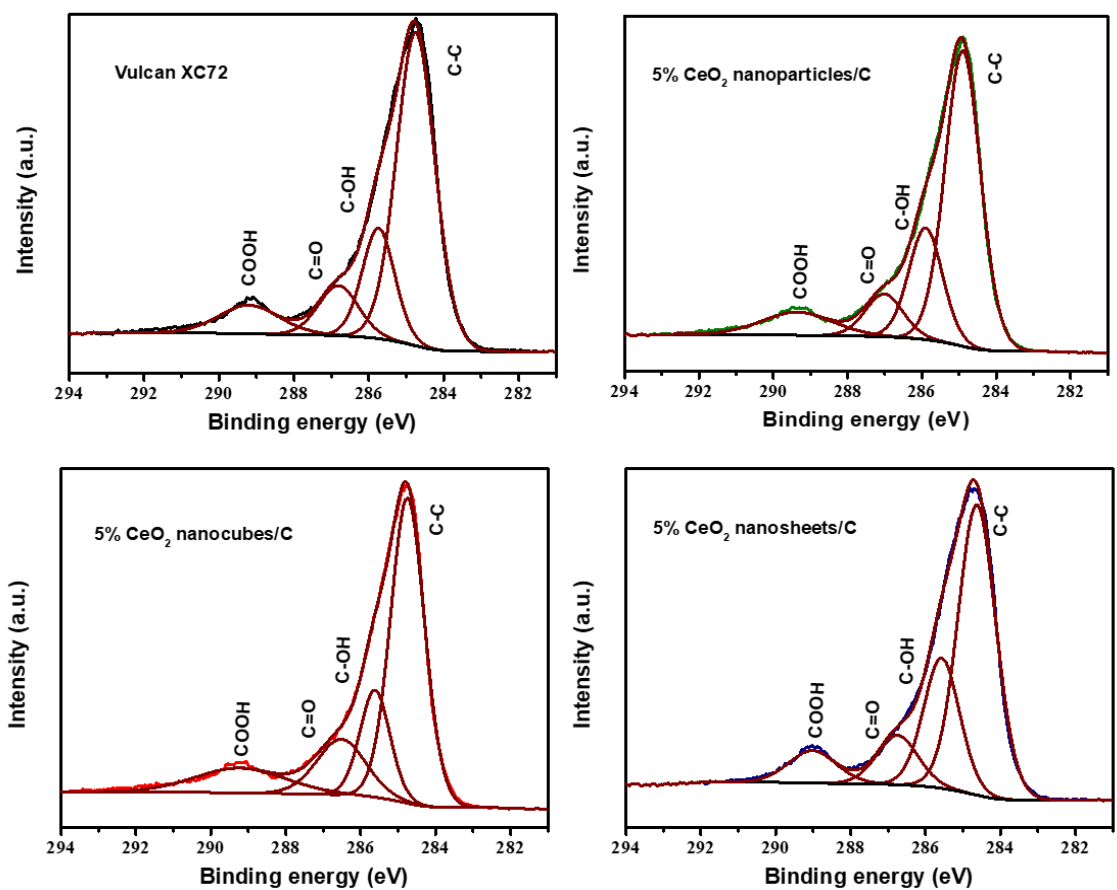

**Figure 4.** Deconvoluted C 1s XPS spectra of Vulcan XC72, 5% $CeO_2$ NPs/C, 5% $CeO_2$ NCs/C, and 5% $CeO_2$ NSs/C.

Vulcan XC72, 5% $CeO_2$ NPs/C, 5% $CeO_2$ NCs/C, and 5% $CeO_2$ NSs/C samples showed 62.25, 59.98, 55.84, and 57.14 at% C-C groups. The oxygen-containing functional groups totaled 37.75 at% for the Vulcan XC72 sample and 40.02, 44.16, 42.86 at.% for the samples 5% $CeO_2$ NPs/C, 5% $CeO_2$ NCs/C, and 5% $CeO_2$ NSs/C.

These findings indicate that all Vulcan XC72 samples, when modified with $CeO_2$, exhibited a higher abundance of surface oxygen-containing functional groups than unmodified Vulcan XC72. Among the $CeO_2$-modified samples, the 5% $CeO_2$ nanocubes/C sample displayed the most substantial concentration of oxygen-containing functional groups. These oxygen-rich functional groups on the electrocatalyst's surface can enhance its hydrophilicity, improving the mass transfer between oxygen and the catalyst surface. Consequently, electrocatalysts with surfaces rich in oxygen functional groups may exhibit higher efficiency in the electrogeneration of $H_2O_2$ [35, 36].

Also, Fig. 5(a-d) displays the deconvoluted O 1s spectra for the Vulcan XC72 samples, 5% CeO2 nanoparticles/C, 5% CeO2 nanocubes/C, and 5% CeO2 nanosheets/C. The O 1s spectra exhibited three main components located at approximately 532, 533, and 534 eV which are related to C=O, -COOH, and water molecules, respectively. The C=O and -COOH groups were also previously identified in the C spectrum, demonstrating coherence between the results found. The 5% CeO2 nanoparticles/C, 5% CeO2 nanocubes/C, and 5% CeO2 nanosheets/C electrocatalysts exhibited an additional low binding energy component (530 eV) associated with O-Ce bonds.

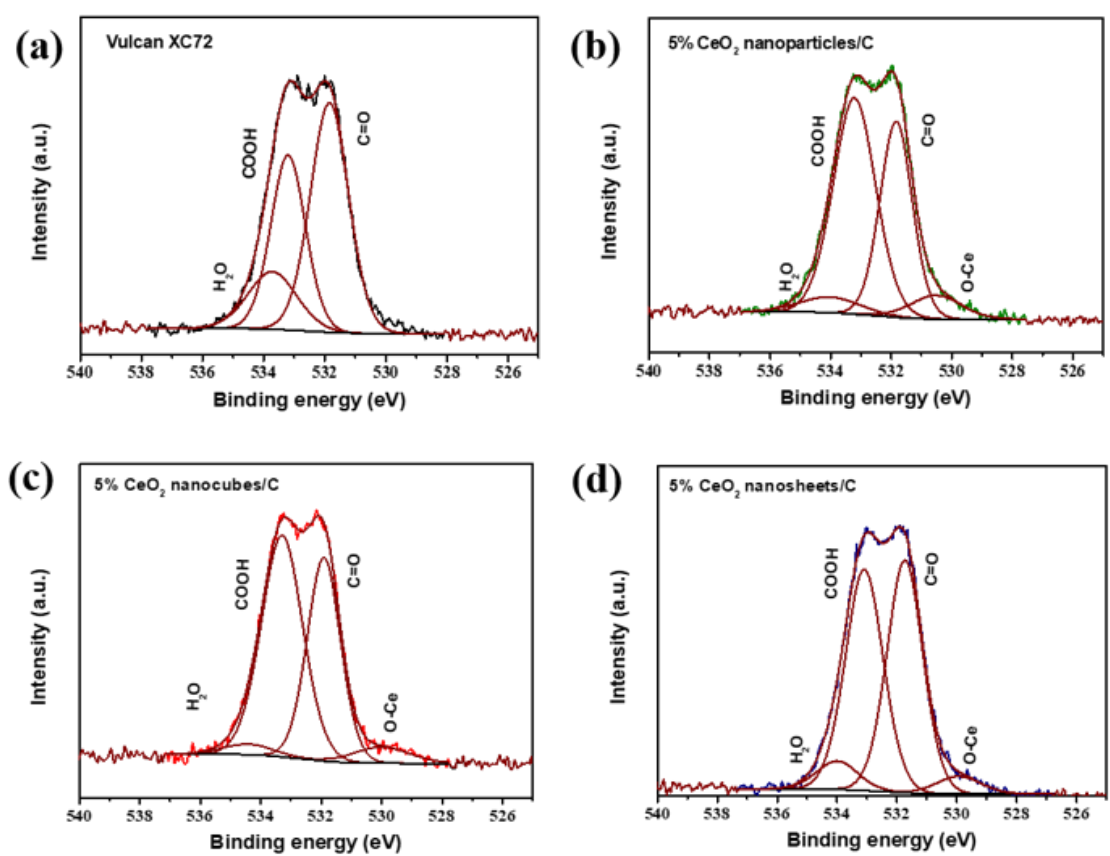

**Figure 5.** Deconvoluted O 1s XPS spectra of Vulcan XC72 (a), 5% $CeO_2$ NPs/C (b), 5% $CeO_2$ NCs/C (c), and 5% $CeO_2$ NSs/C (d).

The concentrations of the C=O and -COOH groups in the Vulcan XC72 electrocatalysts, 5% $CeO_2$ NPs/C, 5% $CeO_2$ NCs/C, and 5% $CeO_2$ NSs/C were quantified at 84.13, 87.86, 91.77, and 89.50% at., respectively. Furthermore, the O-Ce group presents concentrations of 6.76, 5.15, and 4.19 at.% in the electrocatalysts 5% 5% $CeO_2$ NPs/C, 5% $CeO_2$ NCs/C, and 5% $CeO_2$ NSs/C, respectively. These findings are in line with the results discussed in the C spectrum, highlighting the presence of more oxygenated species in the 5% CeO2 NCs/C electrocatalyst.

The contact angle experiment results are presented in Fig. S4, and the corresponding data is summarized in Table 1. Pure carbon Vulcan XC72 exhibited the highest contact angle value at 36.1°. A consistent reduction in contact angle values was observed upon the incorporation of cerium oxide nanostructures. Specifically, the contact angle decreased to 32.8° for $CeO_2$ NPs/C, 29.1° for $CeO_2$ NSs/C, and 26.5° for $CeO_2$ NCs/C. This decline in contact angles can be attributed to the increased presence of oxygenated acid species groups in the modified materials, as corroborated by the X-ray photoelectron spectroscopy (XPS) results following the same sequence, facilitating the transport and adsorption of $O_2$ and a providing better wettability of the electrocatalysts [37, 38].

The modification of the carbonaceous matrix with ceria resulted in enhanced hydrophilicity and wettability of the electrocatalysts. This modification reduced the contact angles, improved active adsorption sites, and facilitated oxygen mass transport. These findings indicate a significant enhancement in the overall performance of the modified electrocatalysts, making them promising candidates for the 2-electron oxygen reduction reaction [39, 40].

**Table 1.** Average contact angle for the unmodified and modified electrocatalysts.

| Electrocatalyst | Contact angle (degree) |
|---|---|
| **Vulcan XC72** | 36.1 ± 3.0 |
| **5% CeO$_2$ NPs/C** | 32.8 ± 2.4 |
| **5% CeO$_2$ NSs/C** | 29.1 ± 1.6 |
| **5% CeO$_2$ NCs/C** | 26.5 ± 2.1 |

### 3.2. Oxygen Reduction Reaction (ORR)

ORR and electrocatalytic activity and selectivity analyses in an alkaline medium (1 mol L$^{-1}$ NaOH) were carried out by linear sweep voltammetry (LSV) using a rotating ring-disk electrode (RRDE) using an $O_2$-saturated electrolyte. Besides the synthetized nanostructures, pure Vulcan XC-72 and commercial Pt/C were also used as references for the 2-electron and 4-electron ORR mechanisms. The steady-state polarization ring and disk curves obtained in those measurements (RRDE 1600 rpm and 5 mV s$^{-1}$) are shown in Fig. 6.

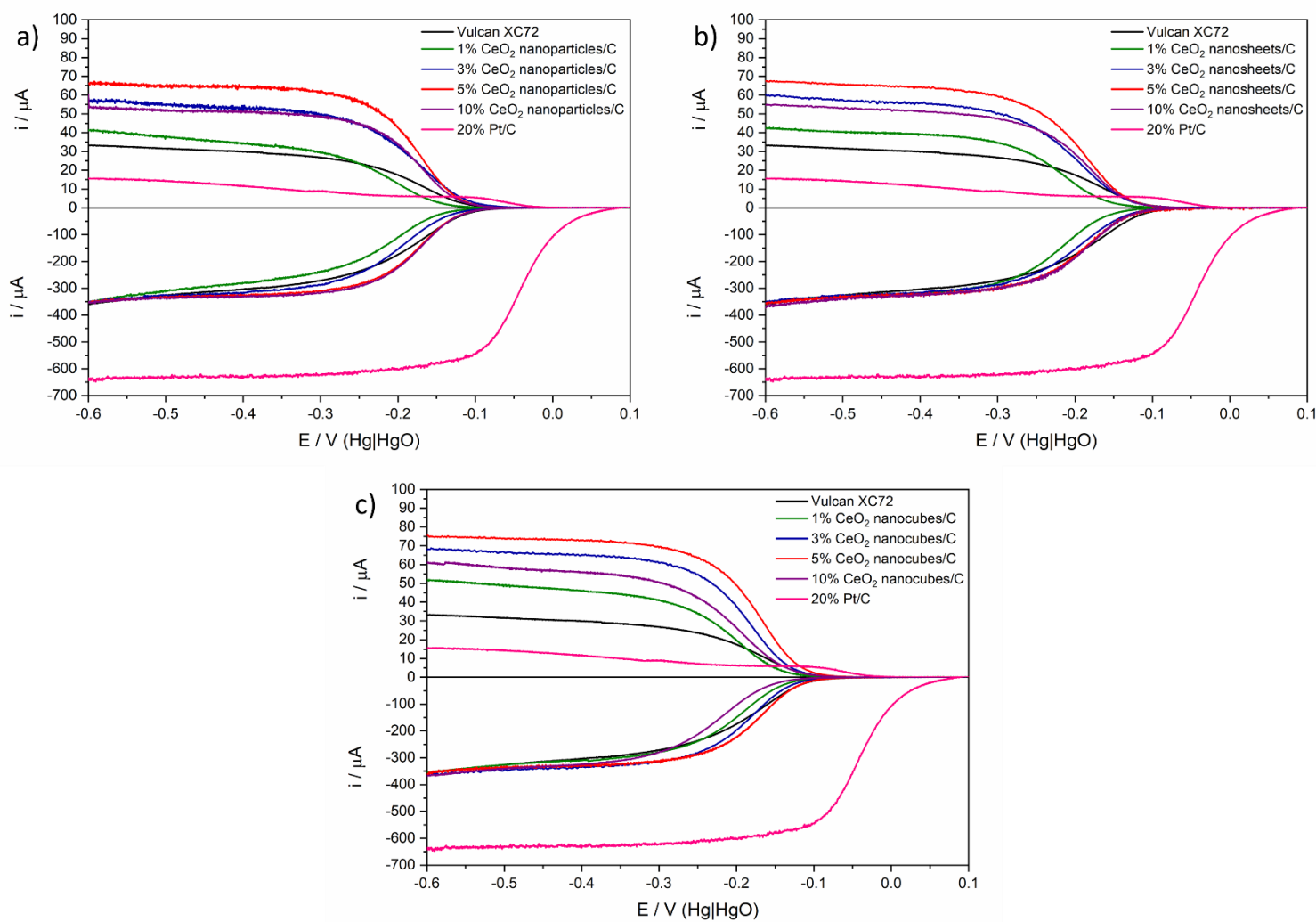

**Figure 6.** Steady-state polarization curves for the ORR for a) CeO$_2$ NPs/C, b) CeO$_2$ NSs/C, and c) CeO$_2$ NCs/C, with the reference materials Vulcan XC-72 and commercial Pt/C in the absence of a magnetic field (ring current $E_{ring}$ = 0.3 V and negative disk current sweep).

The Linear Sweep Voltammetry (LSV) technique involves the electro-reduction of an O$_2$-saturated solution at the disc electrode coated with the electrocatalyst. The resulting electrogenerated H$_2$O$_2$ is transported to the ring electrode through rotation, where it subsequently undergoes oxidation. Notably, elevated oxidation currents observed at the ring electrode correspond to heightened selectivity of the electrocatalyst in promoting the Oxygen Reduction Reaction (ORR) for the generation of hydrogen peroxide [41].

The data depicted in Fig. 5 clearly illustrates that each of the three distinct $CeO_2$ nanostructures remarkably augmented the ring current compared to the pure Vulcan XC-72. In all three instances, the introduction of a mere 1% (w/w) of ceria led to an elevated ring current (32.1 vs. 41.2, 42.4, and 51.4 µA for nanoparticles, nanosheets, and nanocubes, respectively), albeit accompanied by a more negative onset potential for the electrochemical processes.

Incorporating all three cerium oxide nanostructures at a concentration of 3% (w/w) leads to a notable enhancement in ring current during the oxygen reduction reaction. In the case of $CeO_2$ NPs/C, a considerable increase of approximately 78% is evidenced, demonstrating a rise from 32.1 to 57.1 µA. Similarly, implementing $CeO_2$ NSs/C yields a more favorable performance, showcasing an escalation from 32.1 to 60.5 µA, corresponding to an improvement of 88%. Remarkably, the $CeO_2$ NCs/C exhibit the most promising results, elevating the ring current from the baseline of 32.1 µA (unmodified Carbon Vulcan XC-72) to 68.2 µA, reflecting an astonishing gain of 112%.

In light of these remarkable findings, the most striking outcomes were obtained through the incorporation of ceria nanostructures at a concentration of 5% (w/w), as vividly illustrated in Fig. 6. Among these configurations, the $CeO_2$ NPs/C exhibited an astonishing ring current of 66.2 µA—surpassing the performance of pure Carbon Vulcan XC-72 (32.1 µA) by more than twofold. Equally noteworthy, the implementation of $CeO_2$ NSs/C yielded an even more impressive result, elevating the current from 32.1 to 68.4 µA, translating to a remarkable 2.13-fold increase. Notably, the pinnacle of achievement was once again observed with the utilization of $CeO_2$ NCs/C, which yielded an unparalleled ring current of 75.5 µA, showcasing an extraordinary 2.35-fold enhancement.

Incorporating $CeO_2$ nanostructures into Carbon Vulcan XC-72 at a concentration of 10% (w/w) yields a favorable outcome, accompanied by a notable increase in ring current compared to pure Carbon. However, it is worth noting that these results, while promising, fall short of the remarkable outcomes observed with the incorporation of 3% and 5% (w/w) $CeO_2$ NP, NS, and NC. This suggests a concentration-dependent relationship, where the optimal enhancement of the ring current is achieved at lower levels of ceria incorporation. This occurs because the high concentration of nanoparticles, specifically at 10% (w/w) in this study, hinders the active carbon surface area, which is truly responsible for the ORR, as shown by other authors [16, 42, 43].

It is also noteworthy that the addition of all three $CeO_2$ nanostructures to Carbon Vulcan XC-72 at concentrations of 3, 5, and 10% (w/w) distinctly demonstrates a significant improvement in the onset potential of Oxygen Reduction Reaction, shifting it towards more positive values, facilitating the occurrence of those electrochemical process [44, 45].

The quantification of generated hydrogen peroxide and water percentages, along with the electron transferred number (ne-) determination, was accomplished by analyzing polarization curves acquired at 1600 rpm, employing eq. 1-3. In these equations, $i_r$ denotes the ring current, $i_d$ signifies the disk current, and N represents the current collection efficiency of the Pt ring (with N = 0.26). The outcomes of these calculations are visually presented in Fig. 7 [46].

$$H_2O_2\% = \frac{\frac{200 i_r}{N}}{i_d + \frac{i_r}{N}} \quad (1)$$

$$H_2O\% = 100 - H_2O_2\% \quad (2)$$

$$n_{e-} = \frac{4 i_d}{i_d + \frac{i_r}{N}} \quad (3)$$

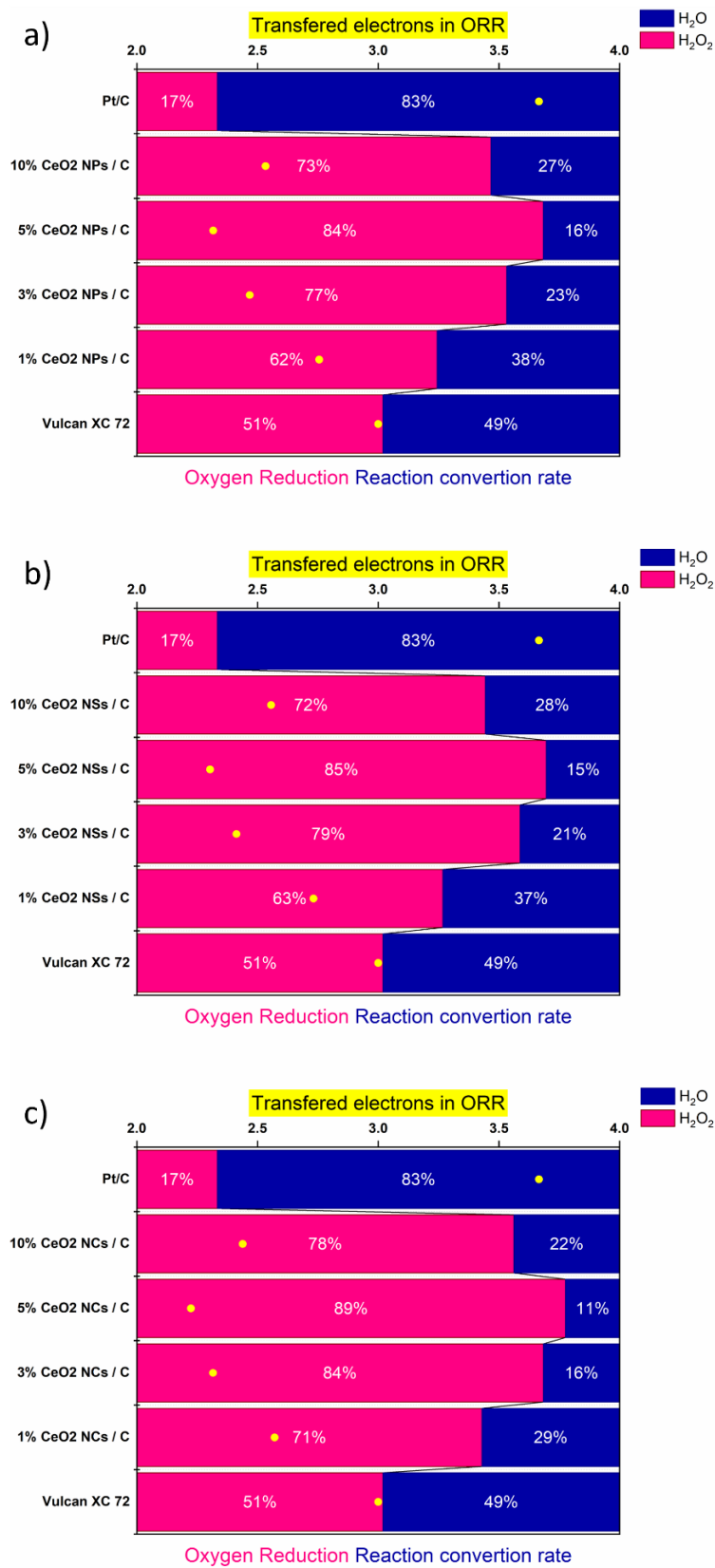

**Figure 7.** Electron transferred number and $O_2$ conversion rate on $H_2O_2/H_2O$ comparison among the different $CeO_2/C$ electrocatalyst nanostructures, as well as Vulcan XC-72 and Pt/C, within a $O_2$-saturated 1 mol L$^{-1}$ NaOH solution in the absence of a magnetic field at -0.6 V.

The results presented in Fig. 7 for the reference materials (bare carbon Vulcan XC-72 and 20% Pt/C) agree with the literature, with the first one favoring the 2-electron ORR with $n_{e^-}$ = 3.0 and the second one leaning the 4-electron ORR with $n_{e^-}$ = 3.7. Applying those materials as electrocatalysts leads to a percentage of $H_2O_2$ of 51% and 13%, respectively [47].

It is also clearly seen from Fig. 7 that adding all tested concentrations of ceria in the carbonaceous matrix favored the 2-electron ORR, with $n_{e^-}$ being close to 2. The percentage of hydrogen peroxide electrogeneration significantly increased in those cases. It is interesting to observe that no matter the structure of the $CeO_2$ nanostructures, they followed the same behavior. The increase in concentration from 1 to 3 to 5% (w/w) in the Carbon Vulcan XC-72 matrix led to a better selectivity in $H_2O_2$ production, going from 62 to 77 to 84% for $CeO_2$ NPs/C, 63 to 79 to 85% for $CeO_2$ NSs/C, and 71 to 84 to 89% for $CeO_2$ NCs/C.

The addition of a 10% (w/w) concentration of the ceria nanostructures resulted in a significant drop in hydrogen peroxide electrogeneration, depicting that carbon matrices modified with smaller amounts of metal oxides are more efficient for the 2-electron ORR, even though all three results are better than bare Carbon Vulcan XC-72.

A table displaying the hydrogen peroxide selectivity for all tested materials in the absence of a magnetic field in different potentials (-0.2, -0.3, -0.4, -0.5, and -0.6 V) is shown in the supplementary material (Table S1).

The results obtained in the RRDE experiments also corroborate the literature regarding the importance of morphological control during synthesis and exposing preferential planes (100) and (110) for ceria nanomaterials, as the $CeO_2$ nanocubes with the preferential exposed plane (100) [31] presents a lower overpotential for the 2-electron ORR, theoretically demonstrated by Lucchetti *et al*. [48, 49].

In summary, at lower $CeO_2$ loadings (1% and 3% w/w), these nanoparticles serve as active sites for the 2e-ORR reaction. $CeO_2$ possesses unique catalytic properties, including high oxygen storage capacity and redox activity. These properties enable $CeO_2$ to facilitate the oxygen reduction reaction, particularly the 2-electron pathway. At these lower loadings, ceria effectively enhances the catalytic activity of the carbon Vulcan XC72 by providing additional active sites and facilitating oxygen adsorption and activation. The presence of $CeO_2$ nanoparticles at moderate loadings (5% w/w) leads to further enhancement of the 2e-ORR performance. At this loading, there is an optimal balance between the active $CeO_2$ sites and the carbon support. The synergistic interaction promotes efficient electron transfer and oxygen reduction, resulting in improved catalytic activity. Additionally, the dispersion of $CeO_2$ nanostructures on the carbon surface may be more uniform at this loading, maximizing the utilization of the active sites. However, when the $CeO_2$ loading is 10% w/w, the synergistic effects between $CeO_2$ and carbon may diminish. Excessive loading can lead to agglomeration or clustering of the nanoparticles on the carbon surface, reducing the accessibility of active sites and inhibiting efficient electron transfer. Additionally, high $CeO_2$ loading may alter the porosity and surface morphology of the catalyst, affecting the diffusion of reactants and products during the electrochemical reaction. These factors collectively contribute to a decline in 2e-ORR performance.

In pursuit of novel approaches to enhance the 2-electron Oxygen Reduction Reaction (ORR), this study repeated the same RRDE experiments aforementioned but this time introducing a pioneering paradigm by incorporating a magnetic field (2000 Oe) generated by a magnet within the electrochemical cell. The application of magnetic fields in electrochemistry represents a recent and innovative development, exhibiting promising prospects for augmenting various electrochemical reactions. The few works in the literature regarding ORR are focused on the 4-electron reaction [50-61]. This novel approach not only expands the horizons of 2-electron ORR research but also underscores the potential of magnetism as a catalyst for transformative advancements in electrochemical science.

With that in mind, the steady-state polarization ring and disk curves obtained in the RRDE measurements (1600 rpm and 5 mV s$^{-1}$) in the presence of a magnetic field are shown in Fig. 8. Also, the quantification of generated hydrogen peroxide and water percentages is illustrated in Fig. 9.

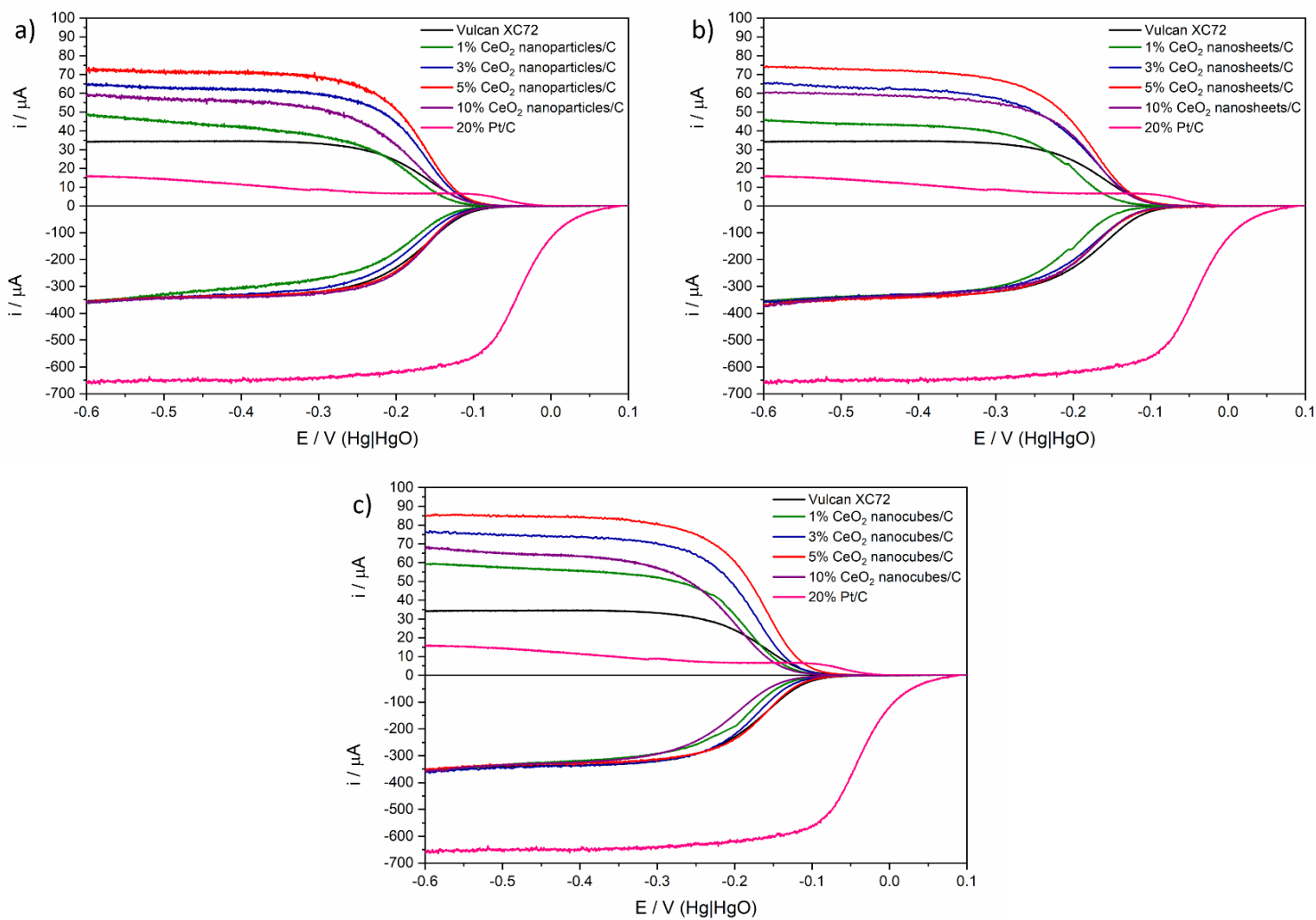

**Figure 8.** Steady state polarization curves for the ORR a) $CeO_2$ NPs/C, b) $CeO_2$ NSs/C, and c) $CeO_2$ NCs/C, with the reference materials Vulcan XC-72 and commercial Pt/C in the presence of a magnetic field (2000 Oe; ring current $E_{ring}$ = 0.3 V and negative disk current sweep).

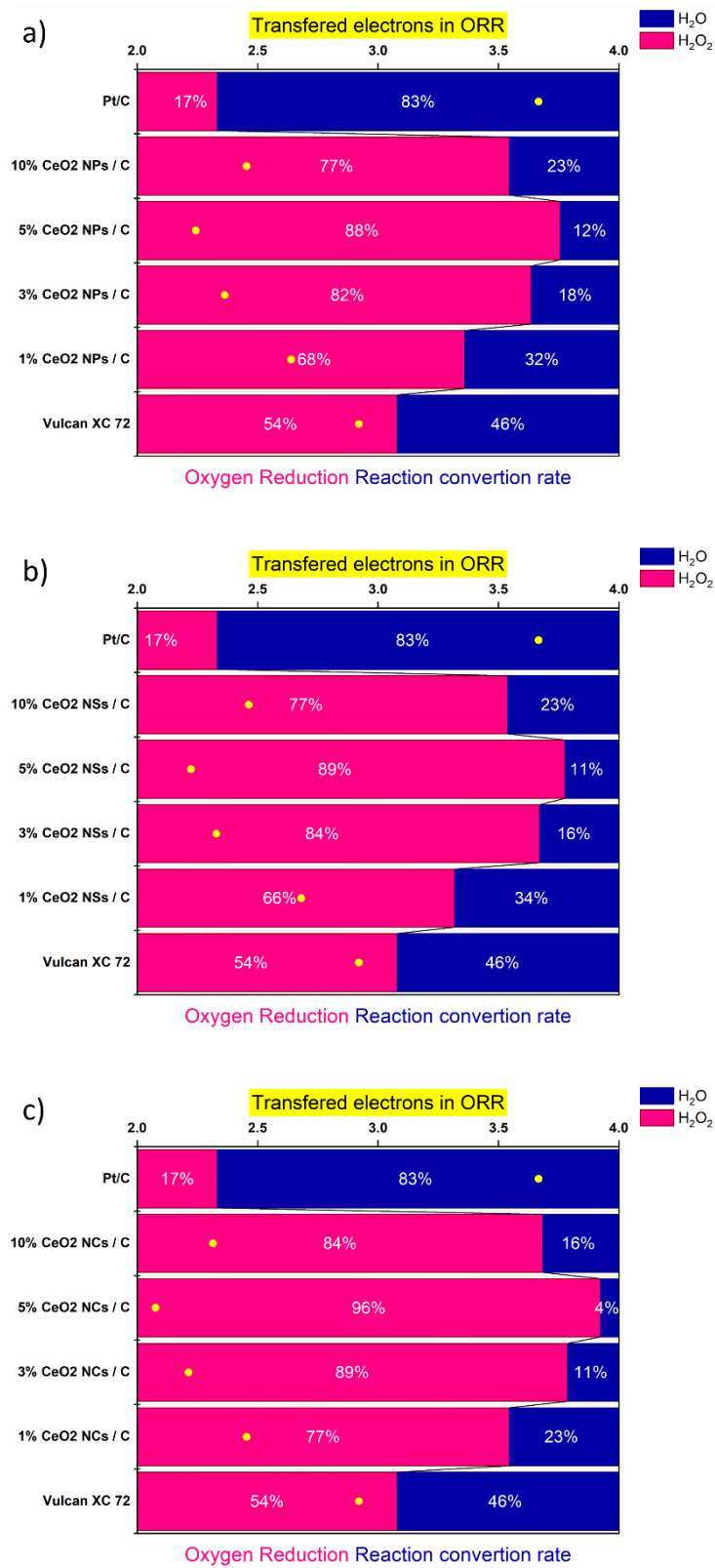

**Figure 9.** Electron transferred number and $O_2$ conversion rate on $H_2O_2/H_2O$ comparison among the different $CeO_2/C$ electrocatalyst nanostructures, as well as Vulcan XC-72 and Pt/C, within a $O_2$-saturated 1 mol $L^{-1}$ NaOH solution in the presence of a magnetic field (2000 Oe) at -0.6 V.

From Fig. 8, although very slightly, there is an alteration in the ring current value for the Carbon Vulcan XC-72 in the presence of the magnetic field (32.1 vs 34.1 µA). This fact is not observed for the commercial 20% Pt/C (15.2 vs 15.5 µA). The theory is that since $O_2$ is a paramagnetic molecule with high magnetic susceptibility, it suffers the effect of the magnetic field. This leads to the Zeeman effect, described as the interplay between the magnetic moment of a lone electron and an external magnetic field, resulting in the division of energy levels about the unpaired electron. The variation observed in states featuring multiple orientations of angular momentum vectors can be attributed to factors such as quantum mechanical spin, classical orbital motion, spin-orbit coupling, or a hybrid combination of these three mechanisms [62-65].

This is reflected in the selectivity of the hydrogen peroxide electrogeneration (Fig. 9), which improved for the Carbon Vulcan XC-72 and maintained the same for the 20% Pt/C in the presence of the magnetic field. Hence, the effect of the magnetic field on the oxygen molecules favors the Pauling adsorption model, favoring the electrocatalyst that is most efficient and selective towards the 2-electron ORR [66-68].

In Fig. 8, it is evident that the presence of a magnetic field exerts a remarkable influence on the ring current exhibited by the three distinct ceria nanostructures across a range of tested loadings while, at the same time, not entirely affecting the disc current. This occurs because it is impossible to distinguish the current for 2 or 4 electron pathways on the disc, but solely in the ring. Notably, enhancements were observed for the $CeO_2$ NPs/C at levels of 11.7%, 12.4%, 10.0%, and 11.3% for concentrations of 1%, 3%, 5%, and 10% (w/w), respectively. Similarly, for the $CeO_2$ NSs/C, increases of 9.6%, 10.2%, 9.1%, and 10.6% were documented. Lastly, concerning the $CeO_2$ NCs/C, the intensifications observed were 15.2%, 11.9%, 12.3%, and 10.8%. These enhancements exceeded the 6.2% followed for the unmodified Carbon Vulcan XC-72. These findings underscore the pivotal role of ceria nanostructures in shaping the observed outcomes.

The highest increase in ring current using the magnetic field towards peroxide electrogeneration was for $CeO_2$ NCs/C 5% Figure 8c, as mentioned without applying the magnetic field. In this case, are three effects: 1) the enhancement of hydrogen peroxide production caused by $CeO_2$ NCs/C 5%, (100) planes are more preferential planes for ceria nanomaterials toward hydrogen peroxide electrogeneration, 2) 5% is the best percentage of cerium oxide nanostructures on Carbon for hydrogen peroxide electrogeneration and, finally 3) the effect of the magnetic field is the best for 1 and 2 together. The selectivity for hydrogen peroxide is almost 100%, which is very impressive. Besides, the potential for oxygen reduction reaction is less negative, indicating both less energy and a low cost of electrogenerated hydrogen peroxide.

A table displaying the hydrogen peroxide selectivity for all tested materials in the presence of a 2000-Oe magnetic field in different potentials (-0.2, -0.3, -0.4, -0.5, and -0.6 V) is shown in the supplementary material (Table S2).

Besides the Zeeman effect in this discussion section, there is also the Kelvin force ($F_K$). It is defined by:

$$F_K = {1}/{2\mu_o} \times c\chi_m \nabla B^2$$

In the context, $\mu_o$ represents the permeability of free space, $\nabla B$ signifies the gradient of the magnetic induction. At the same time, $\chi_m$ and c are the molar magnetic susceptibility and the concentration of the electroactive species in bulk, respectively. The Kelvin force mechanism facilitates the expedited mass transfer of paramagnetic substances, thereby augmenting the reaction kinetics near the electrode. Furthermore, it exerts a discernible impact on the movement of paramagnetic gases, directing the $O_2$ flow towards regions characterized by an intensified magnetic field strength [69, 70].

There is also the contribution of the Lorentz force ($F_L$), generating magnetohydrodynamic effects. The macro- and microscopic convection arose from the combination of the magnetic field and the local current density. Although it is more present in gas-escaping electrodes, in RRDE, the $F_L$ can reduce the concentration polarization. In other words, the Butler−Volmer equation, a key concept in electrochemistry, describes the kinetics of electrochemical reactions, including the ORR. Additionally, Fick's first law, which explains the diffusion of species in a solution, is relevant. It states that diffusion rate is proportional to concentration gradient. The diffusion layer near the electrode surface, where reactants are consumed and products are generated, affects the limiting current density. Under the influence of the magnetic field, magnetohydrodynamic effects induce a flow that reduces the diffusion layer's thickness, enhancing mass transport of reactants to the electrode surface. Consequently, this leads to the observed increase in current density during the 2e-ORR, aligning with predictions from the Butler−Volmer equation and principles of mass transport described by Fick's first law [71, 72].

Regarding the ceria nanostructures, oxygen vacancies, confirmed by the magnetization curves (Fig. 2), can act as some oxygen buffer, as reported in the literature [43]. Cerium oxide exhibits susceptibility to magnetic fields due to ferromagnetic ordering. These orderings originate from unpaired electrons of $Ce^{3+}$ and the existence of oxygen vacancies. In the presence of a magnetic field, an energy level split occurs, as the spin interaction between $Ce^{3+}$ and oxygen vacancies in divergent directions have diverse energies, giving origin to polarons and the so-called bound magnetic polarons (BMP). BMPs occur when a cluster of spins is generated within the Bohr radius. These spins are organized ferromagnetic due to their exchange interaction with an effective mass carrier situated in a localized state [73, 74]. Therefore, the magnetic field boosts the energy states of electron spins within cerium oxide. Consequently, this enhancement facilitates the catalytic reaction.

Building upon this understanding and invoking classical transition state theory, it becomes evident that the magnetic field influences the kinetics of redox electron transfer. Precisely, it elevates the energy states of electrons, subsequently impacting the electron transfer process. As the magnetic field affects the electrochemical reaction, the associated increase in ΔE leads to a tendency for a significantly reduced overpotential, as noted in Fig. S5. This theory was first described by Y. Li *et al.* [75] regarding cobalt oxide nanoparticles for oxygen evolution reaction and was adjusted to our system.

The presence of a 2000 Oe magnetic field introduced several discernible effects that significantly contributed to an increased selectivity for the two-electron mechanism in the Oxygen Reduction Reaction (ORR), as visually depicted in Fig. 9. In the case of 5% (w/w) $CeO_2$ NPs/C, we observed a notable shift in selectivity, rising from 84% to 88%. Likewise, with 5% (w/w) $CeO_2$ NSs/C, selectivity exhibited a marked enhancement from 85% to 89%. Applying a 2000 Oe magnetic field on 5% (w/w) $CeO_2$ NCs/C led to a substantial elevation in selectivity, soaring from 89% to an impressive 96%. These results underscore the profound advantages derived from the magnetic field-enhanced Oxygen Reduction Reaction, particularly concerning electrochemical hydrogen peroxide production.

## 4. Conclusions

In conclusion, our study encompassed the synthesis and evaluation of various $CeO_2$ nanostructures (nanocubes, nanosheets, and nanoparticles) in conjunction with carbon Vulcan XC-72, as electrocatalysts for the 2-electron oxygen reduction reaction. The comprehensive characterization involving XRD, TEM/SEM/EDS, XPS, contact angle, and magnetization curves confirmed these materials' morphology, phases, elemental composition, magnetism, and oxygenated species. Our electrochemical investigations yielded compelling results, revealing that across different concentrations (1, 3, 5, and 10% w/w), all three ceria nanostructures substantially enhanced ring currents in RRDE experiments and exhibited heightened selectivity towards $H_2O_2$ production.

It is worth pointing out that the selectivity for hydrogen peroxide is almost 100% using $CeO_2$ nanocubes (NCs) 5% with the magnetic field. Besides, the potential for oxygen reduction reaction is less negative than all obtained using all the electrocatalysts, indicating both less energy and a low cost of electrogenerated hydrogen peroxide.

Notably, this work broke new ground in the emerging Magnetic Field-Enhanced Electrochemistry, as it introduced the novel element of a continuous magnetic field (2000 Oe) into the 2-electron ORR experiments for the first time. Impressively, the outcomes surpassed those achieved without the magnetic field, showcasing augmented ring currents, increased $H_2O_2$ selectivity, and a reduced negative onset potential. These remarkable enhancements in electrocatalytic performance were attributed to the influence of the magnetic field, which was associated with the Zeeman effect in $O_2$ molecules and the interplay of Lorentz and Kelvin forces alongside Bound Magnetic Polarons within the ceria nanostructures. This novel and original research underscores the transformative potential of incorporating an external magnetic field, thus introducing an innovative dimension to traditional electrocatalysis.

**CRediT authorship contribution statement**

**Caio Machado Fernandes:** Investigation, Validation, Data curation, Writing – original draft. **Aila O. Santos:** Investigation, Validation, Data curation. **Vanessa S. Antonin:** Validation, Writing – review & editing. **João Paulo C. Moura:** Validation, Writing – review & editing. **Aline B. Trench:** Data curation, Writing – original draft. **Odivaldo C. Alves:** Writing – original draft and review & editing. **Yutao Xing:** Investigation, Validation, Writing – review & editing. **Júlio César M. Silva:** Conceptualization,

Writing – review & editing. **Mauro C. Santos:** Conceptualization, Writing – review & editing, Supervision.

## Data availability

The raw/processed data required to reproduce these findings cannot be shared at this time due to legal or ethical reasons.

## Declaration of Competing Interest

The authors declare that they have no known competing financial interests or personal relationships that could have appeared to influence the work reported in this paper.

## Acknowledgements

The authors would like to thank Fundação de Amparo à Pesquisa do Estado de São Paulo (F.A.P.E.S.P., 2021/05364-7, 2021/14394-7, 2017/10118-0, and 2022/10484-4) for the financial support. The authors are also grateful for Coordenação de Aperfeiçoamento de Pessoal de Nível Superior (CAPES) and Conselho Nacional de Desenvolvimento Científico e Tecnológico (CNPq) for their support.